# Perbandingan Shell Unix

Oleh : Spits Warnars Harco Leslie Hendric, S.Kom

Dosen Fakultas Teknologi Informasi Universitas Budi Luhur

Email : spits@dosen.bl.ac.id

## Abstraksi

Is it possible for an Information Technology [IT] product to be both mature and state-of-the-art at the same time? In the case of the UNIX system, the answer is an unqualified "Yes." The UNIX system has continued to develop over the past twenty-five years. In millions of installations running on nearly every hardware platform made, the UNIX system has earned its reputation for stability and scalability. Over the years, UNIX system suppliers have steadily assimilated new technologies so that UNIX systems today provide more functionality as any other operating system.

## Pendahuluan

Jika mendengar *UNIX* orang membayangkan sebagai sebuah hal yang rumit dan sekaligus menantang. Rumit karena orang malas untuk belajar sendiri dan tidak banyak yang pakai. Kalaupun terdengar ada yang memakai system operasi ini hanyalah sebatas praktisi, pecinta unix dan pemakai software yang menghargai hak cipta dan tidak mau mengeluarkan biaya pemakaian software.

*) Kemudahan dari Aplikasi Pemrograman UNIX yang ditemukan oleh Spits Warnars Harco Leslie Hendric, S.Kom

Menantang karena kelas untuk mata kuliah ini sangat banyak peminatnya, dari jumlah maksimal kelas yang ditawarkan hampir semua isi kelasnya penuh, walaupun di Fakultas Teknologi Informasi Universitas Budi Luhur merupakan sebuah mata kuliah pilihan. Apalagi jika membaca istilah kata *UNIX* pikiran orang awam akan mengansumsikan dengan istilah unik yang artinya lain dari yang lainnya.

Ibarat seorang gadis molek yang menantang, system operasi unix ini saat ini menjadi alternatif lirikan yang sangat memikat bagi pengguna software yang terbiasa dengan *freeware*, apalagi negara kita adalah biang keladi dan sumber pembajakan CD bajakan, perangkat keras bajakan, dan banyak bajakan-bajakan lainnya yang secara tidak sadar telah kita gunakan produk bajakan tersebut. Saat ini Microsoft sedang gencar-gencarnya merasia penggunaan software Microsoft bajakan atau software Microsoft yang tidak ber-license.

System operasi *unix* mempunyai struktur system sebagai berikut

a. Struktur system yang mengontrol dan berinteraksi secara langsung dengan hardware yang disebut *kernel*

b. *Utilitas*, perintah-perintah yang tersedia pada system operasi *unix* yang bertindak sebagai pemberi perintah kepada *kernel* untuk menjalankan perintah yang sesuai dengan masing-masing perintah *utilitas*

c. *Shell*, merupakan *utilitas* juga yang bertindak sebagai *interpreter* /penterjemah berdasarkan *utilitas* yang diberikan

d. Aplikasi, program-program yang dibuat dengan sejumlah *utilitas* atau bahasa pemrograman tertentu

Shell sebagai sebuah penterjemah dalam perkembangannya mempunyai beberapa macam jenis diantaranya :

| **Nama shell** | **Prompt** | **Nama Lengkap** | **Pembuat** |
|---|---|---|---|
| bash | $ | Bourne Again Shell | Brian Fox |
| csh | % | C Shell | Bill Joy |
| ksh | $ | Korn Shell | David Korn |
| sh | $ | Bourne Shell | Stephen R. Bourne |
| tcsh | > | Tenex C Shell | Ken Greer, Paul Placeway |

Selain dari *shell-shell* diatas masih ada *shell-shell* lainnya. Karena shell juga merupakan sebuah *utilitas* maka anda dapat melihatnya di direktori *\usr\bin* , seperti *shell tclsh*, dan *zsh*.

**Permasalahan**

Karena banyaknya *shell* yang ada pada system operasi unix dan *shell* mempunyai sintaks pemrograman yang berbeda, sehingga diperlukan suatu perbandingan untuk semua shell tersebut agar membantu pengguna system operasi unix. Ada semacam tuntutan agar programmer dan administrator unix harus tahu semua sintaks *shell*, agar memudahkan mereka dalam merawat system.

Contoh perbedaan shell bash dan csh adalah sebagai berikut

```
$ cat  dodol
  echo "Universitas \n Budi Luhur"
```

Andai kita mempunyai sebuah shell script yang berisi perintah untuk mencetak dengan perintah echo, lalu dijalankan dengan shell bash

$ bash dodol

Universitas \n Budi Luhur

Tampak disini bahwa perintah \n dianggap sebagai sebuah tampilan. Namun apabila kita jalankan shell script tersebut dengan csh maka akan tampil sebagai berikut

$ csh dodol

Universitas

Budi Luhur

Terlihat bahwa perintah \n diterjemahkan oleh shell csh sebagai ganti baris dan kursor ke kiri, seperti halnya perintah \n dalam bahasa C.

Karena adanya perbedaan dalam sintaks shell, maka harus diketahui bagi siapapun pengguna system operasi unix untuk memahami perbedaan sintaks shell, hal tersebut seharusnya menjadi sebuah tuntutan bahwa apabila siapapun yang ingin menguasai system operasi unix, maka si pengguna harus tahu sintaks semua shell. Semoga saja tulisan ini dapat membantu anda untuk dapat memahami dan mempelajari perbedaan sintaks shell.

**Pembahasan**

Secara prinsip apabila anda ingin mempelajari system operasi unix tidaklah sulit. Seperti halnya anda pernah menggunakan perintah-perintah internal dan eksternal pada system operasi Dos, maka utilitas yang ada pada system operasi unix disamakan dengan perintah internal dan eksternal pada system operasi Dos. Contoh jika pada system operasi Dos anda ingin melihat file yang terdapat pada sebuah drive, maka gunakan perintah Dir

C:\>Dir

Maka pada system operasi unix untuk melihat file yang terdapat pada sebuah drive digunakan perintah ls

$ ls

Maka tidaklah terlalu sulit bagi anda untuk mempelajari system operasi unix ini, konsep –nya sama namun ada perbedaan dalam perintah dan sintaks.

Pada tabel diatas kita telah melihat bahwa ada beberapa shell yang ada saat ini, dan ada perbedaan sintaks pada beberapa shell. Kalau demikian akan banyak sintaks yang akan kita pelajari, namun tidaklah demikian. Secara garis besar berdasarkan tampilan prompt nya maka shell diatas dapat kita golongkan sebagai berikut :

a. Shell yang mengikuti sintaks bahasa C (C source)

   Ini dikategorikan sebagai C source dengan asumsi bahwa sebagian besar sintaks yang dipakai menyerupai bahasa C, yaitu shell csh, tcsh, tclsh.

   Shell C source mempunyai tampilan prompt yang bukan $, dimana shell csh dan tclsh mempunyai tampilan prompt % dan shell tcsh mempunyai tampilan prompt >

b. Shell yang mengadopsi sintaks bahasa C ( Not C source)

   Ini dikategorikan sebagai Not C source dengan asumsi bahwa sebagian besar sintaks yang dipakai tidak menyerupai bahasa C, yaitu:  shell bash, ksh, sh, zsh.

   Shell Not C source mempunyai tampilan prompt yang sama yaitu : $

Berikut ini adalah tabel yang mengambarkan perbedaan sintaks pada beberapa shell yang terbagi menjadi 2 golongan yaitu :

| No | Shell bash, ksh, sh, zsh (not C source) | Shell csh, tcsh (C source) |
|---|---|---|
| 1 | Deklarasi variabel<br>A1=10 | <br>**Set** A1=10 |
| | Pada shell Not C source deklarasi dilakukan mirip dengan deklarasi pada bahasa pemrograman lain, sedangkan pada C source menggunakan tambahan perintah *set*. | |
| 2 | Input data/variabel<br>***read*** variable | <br>**Set** variable = $< |
| | Pada shell Not C source menggunakan perintah read, sedangkan pada shell C source menggunakan shell C source | |
| 3 | Perintah kondisi **if**<br>if ungkapan<br>then perintah1<br>else perintah2<br>fi | <br>if (ungkapan) then<br>perintah1<br>else peritah2<br>endif |
| | Pada shell Not C source perintah kondisi if diakhiri dengan fi, sedangkan shell C source perintah kondisi if diakhiri endif dan ungkapan if diletakkan didalam tanda kurung ( ). | |
| 4 | Perintah kondisi berganda<br>case nilai in<br>Pola1) perintah1 ;;<br>Pola2) perintah2 ;;<br>*) perintahn ;;<br>esac | <br>switch (nilai)<br>case pola1 :<br>perintah1 ; perintah2<br>breaksw<br>default: |

<table>
<tr><td></td><td></td><td>perintah4<br>breaksw<br>endsw</td></tr>
<tr><td></td><td colspan="2">Pada shell Not C source perintah kondisi berganda menggunakan sintaks case dan ditutup dengan esac, sedangkan pada shell C source menggunakan sintaks switch( ) dan diakhiri dengan endsw</td></tr>
<tr><td>5</td><td>Perintah perulangan for<br>for indekx [in daftar_argumen]<br>do<br>Perintah<br>Done</td><td>foreach indeks [daftar_argumen]<br>Perintah<br>end</td></tr>
<tr><td></td><td colspan="2">Pada shell Not C source perintah perulangan for menggunakan sintaks for indekx [in daftar_argumen] do dan diakhiri done, sedangkan pada shell C source menggunakan sintaks foreach indeks[daftar_argumen] dan diakhiri end</td></tr>
<tr><td>6</td><td>Perintah perulangan while<br>(Mengulang selama kondisi benar)<br>while ungkapan<br>do<br>Perintah<br>Done</td><td>while (ungkapan)<br>Perintah<br>End</td></tr>
<tr><td></td><td colspan="2">Pada shell Not C source perintah perulangan menggunakan sintaks while (ungkapan) do dan diakhiri done, sedangkan pada shell C source menggunakan sintaks while(ungkapan)</td></tr>
</table>

| | | |
|---|---|---|
| | dan diakhiri end. | |
| 7 | Perintah perulangan lainnya<br>(Mengulang selama kondisi salah)<br>until ungkapan<br>do<br>    perintah<br>done | repeat jumlah perintah |
| | Pada shell Not C source perintah perulangan menggunakan sintaks until ungkapan do dan diakhiri done, sedangkan pada shell C source menggunakan sintaks repeat | |
| 8 | Perintah yang tidak ada di C source<br>***export, readonly, select*** | Perintah yang tidak ada di Not C source<br>***Alias*** |
| | Pada shell Not C source ada beberapa statement tambahan seperti export yang berguna untuk mewarisi variabel dari shell parent ke shell child. Statement readonly digunakan membuat sebuah variabel hanya dapat dibaca saja dan tidak dapat dihapus. Sedangkan statement select hamper mirip dengan perintah pilihan berganda.<br>Pada shell C source ada statement alias yang digunakan untuk memberikan nama arti lain kepada sebuah statement | |

Berikut ini adalah penjelasan dan contoh statement dari tabel yang menggambarkan perbedaan sintaks diatas.

- Inherit variable (not C source): **export**

Digunakan untuk mewarisi/inherit variable dari shell parent ke shell child dan tidak sebaliknya, contoh:

spits@penyamun:~$ a1=10; a2=20

```
spits@penyamun:~$ echo $a1; echo $a2
10
20
spits@penyamun:~$ export a2
spits@penyamun:~$ sh
spits@penyamun:~$ ps
  PID TTY          TIME CMD
10322 pts/0    00:00:01 bash
11242 pts/0    00:00:00 sh
11245 pts/0    00:00:00 ps
spits@penyamun:~$ echo $a1; echo $a2

20
spits@penyamun:~$ csh
% echo $a2; echo $a1
20
a1: Undefined variable.
```

- Variabel read only (not C source)

```
spits@penyamun:~$ readonly a2
spits@penyamun:~$ unset a2
bash: unset: a2: cannot unset: readonly variable
```

- Array pada C source

```
$ tcsh
```

```
> set a=(Univer Budi Luhur Fak Tek Informasi)
>echo $a
  Univer Budi Luhur Fak Tek Informasi
> echo $#a
    6
>echo $a[4] $a[2]
    Fak budi
```

- **alias** pada C source

```
$ tcsh
> dir
dir: Command not found.
> alias dir ls -l
> dir
total 18
-rw-r--r--    1 spits    users          53 Mar 14 19:54 bel
-rw-r--r--    1 spits    users         395 Mar 14 20:22 case1
>
```

- **select** pada not C source

bentuk: **select** nama [in pola ..]

**do**

perintah

**done**

Contoh: $ pico select1

```
select nama in "Dodol" "Rujak" "exit" "Rujak";
do
 case $REPLY in
  1) echo "Ini dodol";;
  2|4) echo "Ini rujak";;
  3) echo "thank you"; break ;;
 esac
done
```

```
$ bash select1
1) Dodol
2) Rujak
3) exit
4) Rujak
#?
```

REPLY → variable yang menampung masukan user dengan perintah select

- Membaca data dari keyboard not C source: **read**

spits@penyamun:~$ cat dodol

```
echo masukin nama
read nama
echo namamu : $nama
```

spits@penyamun:~$ ***bash*** dodol

```
masukin nama
deri
```

namamu : deri

- Membaca data dari keyboard C source **set .. =$<**

spits@penyamun:~$ cat dodol

```
echo masukin nama
set nama =$<
echo namamu : $nama
```

spits@penyamun:~$ ***csh*** dodol

masukin nama

deri

namamu : deri

- perintah kondisi **if**
  - Not C source

Bentuk:

**if** ungkapan

**then** perintah1 ; perintah2

**else** perintah3 ; perintah4

**fi**

**if** ungkapan1
**then** perintah1
**elif** ungkapan2
**then** perintah2
**else**
perintah3
**fi**

Contoh:

```
$ pico if1
read a1
if test $a1  -gt 1
  then echo "a1>1"
       echo "var a1= " $a1
elif test $a1 -eq 0
     then echo "var a1=0"
```

```
else
   echo "var a1=1"
fi
$ bash if1
```

- C source

Bentuk:

```
if (ungkapan) then
      perintah1;perintah2
 else perintah2;perintah4
endif
```

```
if (ungkapan1) then
     perintah1
   else if (ungkapan2)  then
            perintah2
   else
          perintah3
endif
```

Contoh:

```
$ pico if2
set  a1 = $<
if ( $a1  >  1) then
   echo "a1>1" ; echo "var a1= " $a1
else if ( $a1 == 0) then
      echo "var a1=0"
else
   echo "var a1=1"
endif
$ csh if2
```

- Perintah kondisi berganda
  - Not C source

Bentuk:

```
case nilai in
            Pola1) perintah1 ;;
```

Pola2**)** perintah2 **;;**

***)** perintahn **;;**

**esac**

Nb: *) menyatakan selain dari kondisi diatasnya

;; tanda titik koma double harus disertakan pada setiap akhir pola atau *

Contoh:$ cat case1

```
clear
echo "Menu Prog Studi"
echo "1. Teknik Informatika"
echo "2/4. System Informasi"
echo "3. Akuntansi Komputer"
echo -n "Pilih (1,2,3) : "
read pilih
case $pilih in
 1) echo "Ka Prog nya Bang Hari"
    echo "Ditanggung lepet !";;
 2|4) echo "Ka Prog nya Mas goen " ;echo "Pasti O.c !" ;;
 3) echo "Ka Prog nya Den Krisna" ;echo "Tung itung !";;
 *) echo "Ente nyasar kamar orang !!";;
esac
```

```
$ bash case1
   Menu Prog Studi
```

1. Teknik Informatika

2. System Informasi

3. Akuntansi Komputer

Pilih (1,2,3) :

- C source

Bentuk:

```
switch (nilai)
  case pola1 :
      perintah1 ; perintah2
      breaksw
  case pola2 :
      perintah3
      breaksw
  default:
      perintah4
      breaksw
endsw
```

Nb: breaksw → berguna untuk mengarahkan eksekusi ke endsw

Default: → selain dari kondisi diatas

Contoh:$ cat case2

```
clear
echo "Menu Prog Studi"
echo "1. Teknik Informatika"
echo "2. System Informasi"
```

```
echo "3. Akuntansi Komputer"
echo -n "Pilih (1,2,3) : "
set pilih=$<
switch($pilih)
 case 1:
    echo "Ka Prog nya Bang Hari"
    echo "Ditanggung lepet !"
    breaksw
 case 2:
    echo "Ka Prog nya Mas goen Pasti O.c !"; breaksw
 case 3 :
    echo "Ka Prog nya Den Krisna   Tung itung !"; breaksw
 default:
    echo "Ente nyasar kamar orang !!"
endsw
```

- Pengulangan for
  - Not C source

    Bentuk: **for** indekx [**in** daftar_argumen]

    **do**

    Perintah

    **done**

    Contoh: $ pico for1

```
for nama in "Joni Lutung" "Kampret" "dodol"
```

```
do
  echo $nama
done
$ bash for1
 Joni Lutung
 Kampret
 dodol
```

- C source

Bentuk: **foreach** indeks **(**daftar_argumen**)**

Perintah

**end**

Contoh: $ pico for2

```
foreach nama (/usr/bin/bi*)
  echo $nama
  echo "`basename $nama` \n" #backquote
end
$ csh for2
  /usr/bin/biff
  biff

  /usr/bin/bigram
  bigram
```

- Pengulangan while (pengulangan selama kondisi benar)
  - Not C source

    Bentuk: **while** ungkapan

    **do**

    Perintah

    **done**

    Contoh: $ pico while1

```
bil=0
while test $bil -lt 3
do
  echo $bil
  bil=`expr $bil + 2` #backquote
done
$ bash while1
   0
   2
```

  - C source

    Bentuk: **while (**ungkapan**)**

    Perintah

    **end**

    Contoh:

```
$ pico while2
@ bil=0 #pemberian nilai numeric
```

```
while ($bil <3)
    echo $bil
    @ bil += 2
end
$ csh while1
  0
  2
```

- Pengulangan repeat until (pengulangan selama kondisi salah)
    - Not C source

      Bentuk: **until** ungkapan

      **do**

      Perintah

      **done**

      Contoh: $ pico until1

```
bil=a
until echo $bil |grep -v "[^0-9]" > /dev/null
do
  echo "Masukkan bil bulat : "
  read bil
done
$ bash until1
Masukkan bil bulat :
a
```

Masukkan bil bulat :

0.5

Masukkan bil bulat :

1

$

- C source

  Bentuk: **repeat** jumlah perintah

  Contoh: $ pico repeat1

  repeat 3 echo "gile"

  $ csh repeat1

  gile

  gile

  gile

**Kesimpulan**

Dari Perbandingan shell diatas yang dikelompokkan kedalam kedua kelompok yaitu kelompok shell Not C source dan kelompok shell C source, dimana kelompok shell Not C source mempunyai tampilan prompt $ dan kelompok shell C source mempunyai tampilan prompt selain $.

**Saran**

Bagi anda yang ingin menguasai system operasi ini, dibutuhkan kerja keras dan pengerjaan latihan-latihan yang rutin dan terpadu. Selain itu bagi anda yang sudah terbiasa dengan perintah internal dan eksternal pada system operasi DOS akan lebih mudah memahami system operasi unix ini. Perbandingan shell ini akan membantu pemahaman para unixmania dalam mempelajari shell-shell yang ada dalam system operasi unix.

## Daftar Pustaka